\documentclass[10pt,conference]{IEEEtran}
\usepackage{xspace}
\usepackage{mdframed}
\usepackage{color, colortbl}
\usepackage{array}
\usepackage{hyperref}
\usepackage{graphicx}
\usepackage{multirow}
\usepackage{soul}
\usepackage{cleveref}
\usepackage{multirow}
\usepackage{booktabs}
\usepackage{graphicx}
\usepackage[table,xcdraw]{xcolor}
\usepackage[switch]{lineno}
\nolinenumbers

\usepackage{fancybox}

\mdfdefinestyle{MyFrame}{
 outerlinewidth=0pt,
 skipabove=0pt,
 skipbelow=0pt,
 innertopmargin=5pt,
 innerbottommargin=0pt,
 linewidth=0pt,
 topline=false,
 rightline=false,
 leftline=false,
 innerrightmargin=4pt,
 innerleftmargin=4pt,
 nobreak=false}

\makeatletter
\author{\IEEEauthorblockN{Elena Lyulina}
\IEEEauthorblockA{JetBrains Research\\
Saint Petersburg, Russia \\
elena.lyulina@jetbrains.com}
\and
\IEEEauthorblockN{Mahmoud Jahanshahi}
\IEEEauthorblockA{Babol, Iran\\
mahmoud.jahanshahi@gmail.com}
}
\newcommand{\linebreakand}{%
  \end{@IEEEauthorhalign}
  \hfill\mbox{}\par
  \mbox{}\hfill\begin{@IEEEauthorhalign}
}
\makeatother

\definecolor{Gray}{gray}{0.9}

\newcolumntype{L}[1]{>{\raggedright\let\newline\\\arraybackslash\hspace{0pt}}m{#1}}

\usepackage{ifthen}
\usepackage{amssymb}
\DeclareOldFontCommand{\sf}{\normalfont\sffamily}{\mathsf}
\newboolean{showcomments}
\setboolean{showcomments}{true}
\ifthenelse{\boolean{showcomments}}
  {\newcommand{\nb}[2]{
    \fcolorbox{Gray}{yellow}{\bfseries\sffamily\scriptsize#1}
    {\sf\small$\blacktriangleright$\textit{#2}$\blacktriangleleft$}
   }
   
  }
  {\newcommand{\nb}[2]{}
   
  }

\begin{document}

\title{Building the Collaboration Graph of Open-Source Software Ecosystem}
\maketitle

\begin{abstract}
    The Open-Source Software community has become the center of attention for many researchers, who are investigating various aspects of collaboration in this extremely large ecosystem.
    Due to its size, it is difficult to grasp whether or not it has structure, and if so, what it may be.
    Our hackathon project aims to facilitate the understanding of the developer collaboration structure and relationships among projects based on the bi-graph of what projects developers contribute to by providing an interactive collaboration graph of this ecosystem, using the data obtained from World of Code~\cite{ma2019} infrastructure.
    Our attempts to visualize the entirety of projects and developers were stymied by the inability of the layout and visualization tools to process the exceedingly large scale of the full graph.
    We used WoC to filter the nodes (developers and projects) and edges (developer contributions to a project) to reduce the scale of the graph that made it amenable to an interactive visualization and published the resulting visualizations.
    We plan to apply hierarchical approaches to be able to incorporate the entire data in the interactive visualizations and also to evaluate the utility of such visualizations for several tasks.
\end{abstract}

\section{Project Goals}
With the rise of projects and people involved in the OSS (Open-Source Software) ecosystem, the collaboration patterns of developer communities are evolving in such a manner which does not allow being thoroughly investigated easily, due to the huge amount of data available.
Being able to recognize these patterns would support researcher in various fields of science to better understand the behavior at the ecosystem level and to utilize this knowledge for achieving their specific research objectives.
Trying to understand evolution patterns of open-source software systems and communities~\cite{nakakoji2003} or focusing on the strategies and processes by which new people join the existing community of software developers~\cite{krogh2003} are some examples of such attempts.

WoC (World of Code) stores the huge and rapidly growing amounts of data throughout the entire OSS ecosystem, and provides basic capabilities to efficiently extract and analyze the data at that scale. Compared to investigating network structure of social coding~\cite{thung2013network} or exploring the patterns of social behavior ~\cite{yu2014exploring} only in GitHub, using WoC unique infrastructure, we aim to build a graph of collaboration across all OSS to address the need for an exploratory analysis at that level. 
Visualization techniques are powerful tools to support such exploration tasks and, more generally, to support information retrieval and classification~\cite{borner2005}.

The primary aim of our hackathon project was to build a visualization in the form of a graph of the collaborations in the entire OSS ecosystem.
We focused on building an easy-to-use exploration tool, so that it could be interactively utilized by researchers. 
Potential use cases of this tool are identifying author communities and collaboration patterns as well as project clusters and common underlying features, among others.

\begin{figure*}[h!]
\centering
\begin{minipage}{.45\textwidth}
  \includegraphics[width=\linewidth]{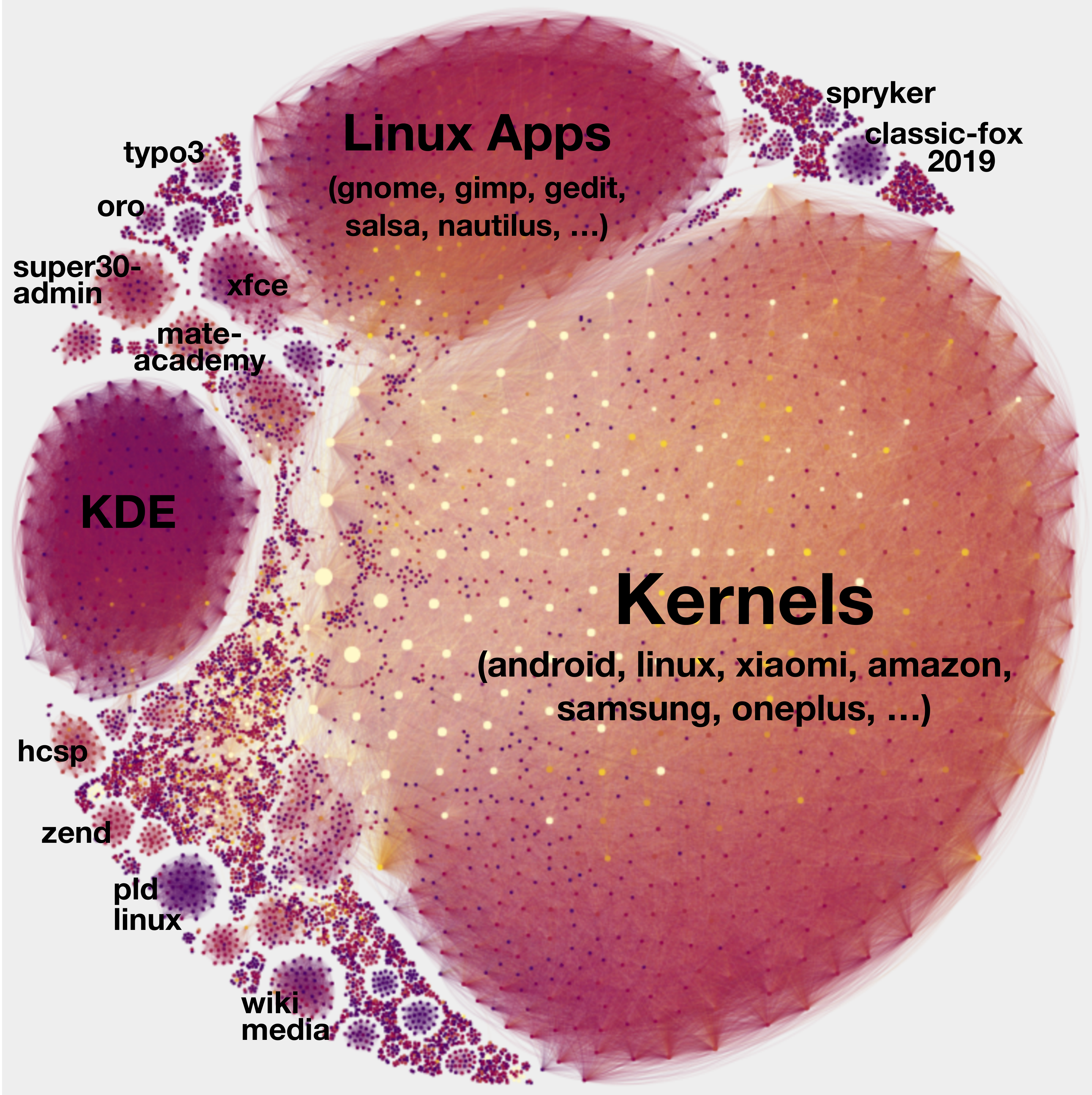}
  \caption{Graph of projects with manually labeled clusters using the project names}
  \label{projectsgraph}
\end{minipage}%
\hspace{.3cm}
\begin{minipage}{.45\textwidth}
  \includegraphics[width=\linewidth]{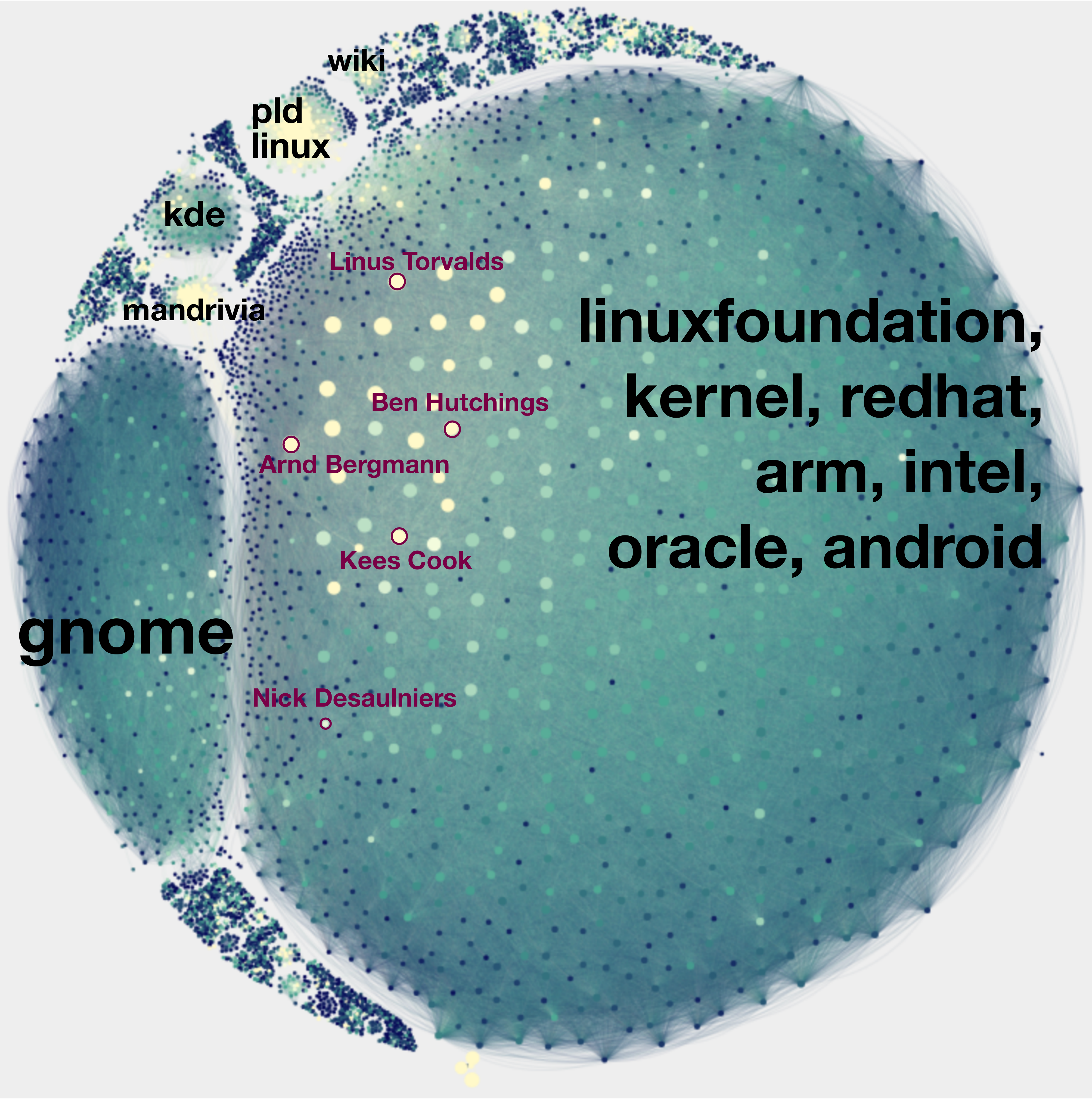}
  \caption{Graph of authors with manually labeled clusters using authors' email domains}
  \label{authorsgraph}
\end{minipage}
\end{figure*}

\section{Approach}
\subsection{Using WoC}
The open source project data that can be obtained through WoC infrastructure provides a unique opportunity to have access to the whole OSS data at the same time which is vital to our goal.
We define collaboration between authors as contribution to the same project based on WoC project-to-author (p2a) maps, which are constructed as a composition of project-to-commit and commit-to-author maps.
For the starting point, we used the projects metadata which are available through WoC MongoDB tables\footnote{version R} with the summary information about authors and projects to enable the selection of subsets for later analysis.
We retrieved the list of projects with at least two authors to exclude single author projects, as they do not contribute any project-based collaboration information. 
It resulted in a total number of 21,978,139 projects.
We then used WoC p2a maps to obtain the list of all author IDs involved in these projects.
This query resulted in 62,067,751 rows representing unique combinations of a project name and an author ID. These 62 million rows contained 25,880,258 unique author IDs.

Investigating our data, we came to the understanding that there were two major flaws in our table.
One being multiple author IDs associated with one author, 
as a single developer may use multiple author IDs in their work; 
and the other one was author IDs that were not clearly indicating a unique author (e.g.,``\textsuperscript{$\wedge$}\textsuperscript{$\wedge$} \textless\textsuperscript{$\wedge$}\textsuperscript{$\wedge$}\textgreater").
To address the first issue, we used WoC author ID aliasing (a2A) maps~\cite{fry2020} through which we were able to aggregate about 11 million author IDs into 5 million unique Authors.
Adding 14 million author IDs which were not aliased in a2A maps, we reached 19 million unique authors in our data.
For the second issue, we filtered out invalid author IDs by using regular expressions\footnote{[A-Za-z0-9.\_\%+-]\{1,\}@[A-Za-z0-9.-]\{1,\}.[A-Za-z]\{2,\}} based on email addresses.
This removed approximately 2 million entries with invalid email addresses from the author list.
The final table\footnote{Can be found in the replication package at \url{https://zenodo.org/record/4626189\#.YFfcQrQvN70}} contained 47,400,507 rows of unique project name and author combinations, containing 21,838,782 unique projects and 17,640,565 unique authors.
For our visualizations we used this link data to depict two graphs:
\begin{itemize}
    \item{Graph of projects as the nodes with the mutual authors between them representing edges};
    \item{Graph of authors as the nodes with the mutual projects between them representing edges}.
\end{itemize}
Technically, these are the so-called, multi-edges, as a single author may link more than one project and one project may link more than two authors.

\subsection{Drawing graphs}

Visualizing this huge amount of data presents many challenges, such as layout and rendering problems, and high computational cost.
However, the biggest problem is to catch the overall structure of the OSS ecosystem.
To prevent the networks looking like hairballs, the data was filtered by excluding graph nodes and edges of insignificant value. The filtering thresholds were adjusted manually for each graph in a try and error manner, so that the visualization reflects the tightly-knit communities.
This resulted in a graph of projects with at least 50 authors each, connected with each other if there were at least 50 mutual authors (Fig.~\ref{projectsgraph}); and a graph of authors with at least 60 projects each, connected with each other if they have had at least 30 common projects (Fig.~\ref{authorsgraph}).
The total number of nodes in each graph are 6514 and 6673 respectively.

The visualization was performed using Gephi~\cite{gephi} with  ForceAtlas2~\cite{forceatlas} layout. The nodes are colored according to the number of projects for authors and the number of authors for projects, where a lighter color represents a higher number, and the size of the nodes correlates with their weighted degree.  
Since the project’s goal was to create an interactive and easy to explore visualization, we built an HTML page for each graph using  SigmaExporter~\cite{sigmaexporter} plugin.

\section{Preliminary results}
The resulting graph pages are available online through the project repository\footnote{\url{https://github.com/woc-hack/collab-graph}}. Each of them depicts an interactive network with features such as zooming in and out, searching by node names, viewing the connections of selected nodes, and may serve as a useful tool to investigate the collaborative patterns of the OSS community. 

For example, using the graph of projects we can easily discover and describe different sub-communities by manually examining project names in clusters, since each of them represents repositories with a significant number of mutual collaborators (see Fig.~\ref{projectsgraph}).
The most notable projects from the largest cluster are various kernels: Android, Linux, Xiaomi, Amazon, Samsung, OnePlus, etc.
Even though the other projects are less well-known, making it difficult to understand their purposes by name, most of them contain the word ``kernel".
Given that fact, the biggest cluster can be described as kernels.
The names from the second largest cluster represent various easily recognizable Linux applications: Gnome, Gimp, Gedit, Salsa, Nautilus, and so on.
Finally, most of the labels from the third largest cluster contain the word ``kde", which suggests that this cluster might represent KDE, an international free software community.

We can also perform the same analysis on the authors' graph and even link some clusters of both graphs. While for some well-known authors (like Linus Torvalds, Ben Hutchings, Arnd Bergmann) we can easily identify the topics of projects they work on, this is generally not the case. However, we can analyze clusters by using authors' email addresses, assuming that many developers use their corporate emails and the domains may indicate their company and subsequently its projects (see Fig.~\ref{authorsgraph}). Using this approach, we can find the most common domains (\texttt{linuxfoundation.org}, \texttt{kernel.org}, \texttt{redhat.com}, ...) from the biggest cluster of the authors' graph, which are clearly related to the topics of the Kernel cluster of the projects' graph. Hence, we can assume we found the group of authors who work on the Kernels group of projects. Similarly, we can find authors of Linux Apps (they mostly use \texttt{gnome.org} domain), KDE (\texttt{kde.org}), PLD Linux (\texttt{pld-linux.org}), Wikimedia (their domains contain keyword ``wiki"), linking the clusters of both graph.

Although this analysis is superficial, it can be done very quickly using our interactive visualization and may support other research into deeper study of OSS social networks. For example, the cluster labeling performed above may serve as the first step of a more thorough cluster analysis using supplemental information about projects (e.g. their description, source code, commit history, etc.).
Besides cluster analysis, such studies may include:
\begin{itemize}
    \item \textit{Finding key people and projects of OSS.}
    Using our interactive visualization, it is easy to highlight candidates for the key nodes as their color and size are likely to correlate with their significance.
    \item \textit{Comparing the structure of various subcommunities.} Collaboration processes in each subcommunity of developers may differ from each other resulting in different structures of connections that can be observed on authors' graph.
    \item \textit{Discovering graph anomalies.} Some nodes of the graphs may look like outliers and are interesting to be studied deeply. For example, we found a big project connected with many various clusters namely \textit{helpmeplz}\footnote{\url{https://github.com/everyonehelp/helpmeplz}}, which is quite unusual. Further investigation revealed that it has thousands of contributors, but only contains one text file and its commit history is full of ``im helping" messages. 
\end{itemize}

\section{Future work}
\subsection{Collaboration graph}
Investigating through the pre-built graph could overwhelm the user with the huge amount of data which they may not necessarily need.
Therefore, it would be useful to support the interactive build of a graph subset based on user selection.
In addition, the current tool can be improved by adding metadata to nodes and edges, which can be taken from the above-mentioned summaries of authors and projects in WoC. In addition, other useful data points such as files, organization and etc. can be used to build graphs, which can be pursued to expand the scope of the tool. 

\subsection{Expansion of WoC}
Being able to build the graph interactively needs certain features added to the WoC infrastructure, for it is time consuming to preprocess the data now. These features could improve the performance of our tool or any other tool for that matter, to retrieve the data from WoC and build the graph as per request of the end user.

Just as a demonstration, we have built a2gr tool which reads an author ID as input, gets depth as input argument and outputs the list of nodes and edges to build the graph around the given author ID with requested depth, defined as the maximum distance between the central node and any given node in the graph.
Having maps linking authors who have mutual projects (a2a) and projects with mutual authors (p2p) or similar maps in WoC, which contain the neighboring entities of each entity may prove very useful in this regard.

\section*{Acknowledgment}
We would like to thank all the MSR 2021 Hackathon track PC members for their valuable comments during the event. Especially, we thank Professor Audris Mockus for his professional guidance and beneficial support during the planning and development of this work.
We would also like to thank Vladimir Kovalenko who proposed the original idea of this work.

\bibliographystyle{IEEEtran.bst}
\bibliography{IEEEabrv, cite.bib}

\end{document}